**Article type: Full Paper**

**Superradiantly limited linewidth in complementary THz metamaterials on Si-membranes**

*Janine Keller[1]\*, Johannes Haase[2], Felice Appugliese[1], Shima Rajabali[1], Zhixin Wang[1], Gian Lorenzo Paravicini-Bagliani[1], Curdin Maissen[1], Giacomo Scalari[1]\*, Jérôme Faist[1]*

J. Keller, F. Appugliese, Z. Wang, S. Rajabali, G. L. Paravicini-Bagliani, Dr. C. Maissen, Dr. G. Scalari, Prof. Dr. J. Faist
Institute for Quantum Electronics, ETH Zürich, 8093, Switzerland
E-mail: janine.keller@phys.ethz.ch, scalari@phys.ethz.ch

Current address of Dr. C. Maissen
Nanooptics Group, CIC nanoGUNE, Donostia - San Sebastian, 20018, Spain

Dr. J. Haase
Paul Scherrer Institute, Villigen, 5232, Switzerland

We study complementary double split ring THz resonators fabricated on a 10 μm thin Si-membrane. The linewidths of the fundamental LC-mode and dipolar mode are drastically narrowing with increased resonator spacing. The extracted decay rate of the LC-mode as a function of the resonator density shows a linear dependence, evidencing a collective superradiant effect of the resonator array. Furthermore, we show that a metamaterial can be designed for a low superradiant broadening of the resonance at high resonator densities, i.e. in the metamaterial condition. The use of a thin membrane as a substrate is crucial, since it shifts the THz surface plasmon polaritons modes to much higher frequencies, preventing them to couple to the LC mode and unveiling the superradiant broadening mechanism for a large range of lattice spacings. At higher frequencies, not interfering with the high Q LC-mode, other additional modes, which we ascribe to photonic crystal modes, form in the Si-membrane. We map the angle dependent band structure and show corresponding simulated electric field distributions.

# 1. Introduction

Metamaterial[1,2] research has prompted the photonics community to investigate the possibilities to design and control electromagnetic properties in transmission, reflection and absorption. The manifold parameters, including the choice of material,[3-5] geometry[2,6,7] and arrangement of the meta-atom in typically two or three dimensions offer a rich palette of attributes.[1,8-12] Achieving a narrow resonance linewidth, remains a desirable feature for many areas, e.g. for sensing,[13] narrow band filters[14,15] or light-matter coupling experiments.[16-19]

One of the most successful structures in meta-atom designs is the so-called (direct) split ring resonator (SRR). A ring type element with an opening forms an inductance (L) with a capacitor (C), which behaves as a LC-circuit. The influence on the resonator frequency and linewidth by varying the arm length and gap openings have been studied in detail[20] as well as various resonator shapes,[6] especially also engineering electric split ring resonators.[6] An electric SRR can be engineered by two combined SRR, thus cancels the magnetic response which results in a pure electric response. The coupling of the meta-atom array elements also plays a major role in the observed resonance spectrum. Single split ring resonators were found to show magneto-inductive and electro-inductive coupling due to the respective magnetic and electric dipoles,[21-23] whereas in double (electrical) SRR, only electro-inductive waves were observed.[24] In the case of direct metamaterials it was additionally found that a diffraction grating effect [14, 25-34] arises when the strict metamaterial condition, i.e. placing the elements with highly subwavelength spacing, is violated. Instead, for complementary SRR (cSRR),[35,36] which can be obtained by exchanging the role of metal and dielectric leading to the same resonance frequency (according to Babinet's principle), we observed the excitation of surface plasmon polaritons (SPPs) for arrays of complementary SRR in our previous work.[37]

In our previous study .[37] we observed that the SPP couples strongly to the LC-resonance when tuning the lattice constant displaying an anti-crossing behavior and showed that the SPP resonance is a substrate supported bound wave. In the present study, we demonstrate that we can control the frequency of the SPP by placing the complementary resonators on a Si-membrane, which is thinner than the decay depth of the SPP modes into the substrate at the investigated frequencies.[38] Therefore, the frequency of a bound SPP in the substrate is detuned far away from the fundamental LC resonance and we can observe the intrinsic influences of the inter meta-atom distance on the fundamental resonances of the array, well into the regime beyond subwavelength spacing between the cSRR. We uncover a linear dependence of the decay rate of the LC-resonance linewidth on the density of resonators. Such linear decay rates are typically linked to the description of a collective behavior of coupled oscillators, as first described by Dicke[39] quantum mechanically for a set of two level systems. Since then, the effect was coined as "superradiance". Subsequently, it has been observed in many different systems, including cyclotron resonances,[40] quantum dots,[41] intersubband transitions[42] and direct metamaterials.[20,22,43] These recent observations include both quantum and classical system, where similar effects are observed when studying the cooperative behavior, as noted already in the work of Rehler and Eberly.[44] Also other cooperative effects stemming from the array size rather than the density of elements have been reported.[45-47] In the work of Wenclawiak and coworkers,[41] the authors present the approach to change the number of direct meta-atoms randomly placed in supercells and show a Dicke-like superradiance. Here, however, we employ the more conventionally used approach to repeat the meta-atoms in a regular array configuration. By using the membrane, we can show not only the superradiance effect but also extend the range of the lattice spacing study well beyond the limit of subwavelength spacing of the resonators. Furthermore, we modified a complementary split ring resonator to limit superradiant broadening at high resonator densities with respect to the standard geometry. We show, that a small capacitor gap with less radiative coupling directly influences the superradiance properties of the array. Placing resonators on thin membranes was investigated for applications like enhanced sensing or filters,[48,49] or for the use of a flexible substrate[50] for direct metamaterials. The additional rejection of broadband background transmission by the use of a complementary SRR has been shown to be advantageous e.g. in light-matter interactions,[51] where also a high quality factor to map new phenomena[52] proved to be of importance. With this study, we show how to obtain a high-density array of complementary meta-atoms with high quality factors and strong electric field enhancement.

## 2. Results and Discussion

We fabricated five arrays of complementary split ring resonators with different lattice constants *a* on 10 µm thin Si-membranes commercially available from Norcada, as sketched in Figure 1. The 10 µm thin Si-membrane windows have dimensions of 4.8 mm x 4.8 mm and are held inside a 300 µm thick Si frame with 10 mm x 10 mm. Each array with a dimension of 4 mm x 4 mm is written via electron beam lithography in the center of a membrane. After development of the resist, we deposit 250 nm of copper followed by 20 nm of gold as capping layer. The lift off is done in hot acetone, supported by ultra-sonication at a frequency of 135 MHz. The resonator dimensions are chosen to be identical to our previous study in which the resonators were placed on 625 µm GaAs substrates instead on thin membranes.[37] The meta-atoms feature small capacitive gaps 500 nm-wide where the electric field is confined and strongly enhanced on a subwavelength scale. A THz probe field that is polarized across the capacitive gap, i.e. along the y-direction in Fig. 1(a), reveals a resonance frequency of $f_{LC}$ = 1.12 THz of the fundamental LC-mode. Terahertz (THz) transmission spectra are measured using a standard THz time-domain spectroscopy setup where the sample is at room temperature. The THz generation with a photoconductive switch[53,54] covers a bandwidth from 0.1 THz to 3 THz. More details can be found in Ref. [37]. The linewidth for the membrane measurements is resolution-limited to 15 GHz.

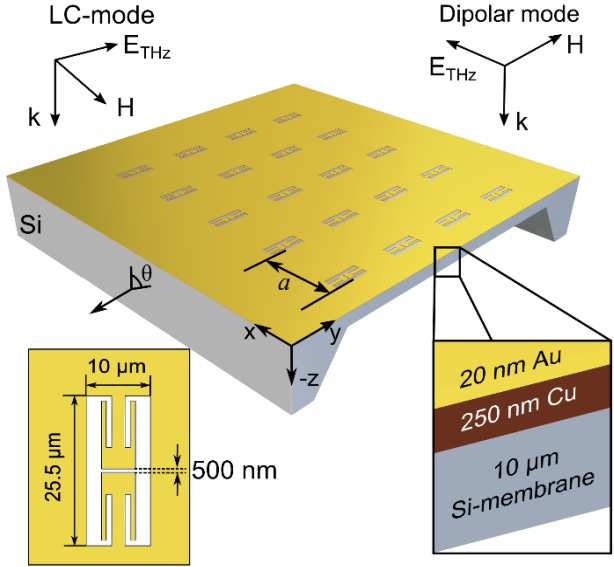

*Figure 1: Sketch of the sample layout (not to scale) with an array of resonators on a 10 µm Si-membrane. Indicated are the electric field polarization direction to excite the LC-mode of the resonator (along the y-axis) and the dipolar mode (along the x-axis). The overall resonator dimensions (10 µm x 25.5 µm) and the 500 nm capacitor gap are indicated in a close up sketch. Further details on all dimensions can be found in our previous work in Ref. [37].*

## 2.1 Linewidth broadening of LC-resonance

The transmission spectra for lattice constants $a$ = 40 µm, 50 µm, 75 µm, 100 µm and 180 µm are displayed in Fig. 2 (a) showing the fundamental LC-mode at $f$ = 1.12 THz, with the transmission of $a$ = 180 µm multiplied by 10 to be visible on the same scale. For better visibility of the linewidth of the different arrays, in Fig. 2 (b) the normalized transmission spectra for the lattice constants $a$ = 40 µm, $a$ = 75 µm and $a$ = 180 µm are displayed. The full widths at half maximum (FWHM) of the resonance features are extracted via Lorentzian line fits and are indicated in the graphs. We observe a narrowing of the LC-resonance with increasing lattice constant, reaching a FWHM value of $\Delta f$ = 63 GHz (fit accuracy: ± 0.5 GHz) at the largest measured lattice constant $a$ = 180 µm. Thus the quality factor, defined as $Q = f/\Delta f$, increases with increasing lattice constant or, in other words, decreasing meta-atom density. It reaches up to a value of $Q = 17.9$ for $a = 180$ μm. The decay rate $\Gamma$ of the resonance is closely related to the quality factor: $Q = f\pi/\Gamma$. By plotting the decay rates of the five measured cSRR arrays as a function of resonator density $\rho = 1/a^2$ times the squared wavelength $\lambda$ (Fig. 2 (b)), we observe a clear linear dependence. We interpret this effect as "superradiance",[39] which is observable in quantum and classical systems.[45] In an illustrative picture, this effect can be described as a collective coupling effect of the electric dipoles of the meta-atoms. When all resonators oscillate in phase and interact with the THz field, the density of resonators leads to a homogeneous broadening of the resonance linewidth, as similarly observed for direct single loop split ring resonators in Ref. [22]. The homogeneous nature of the observed linewidth broadening is reflected clearly in the observed Lorentzian line shape, in contrast to a Gaussian line shape for inhomogeneous broadened spectra.

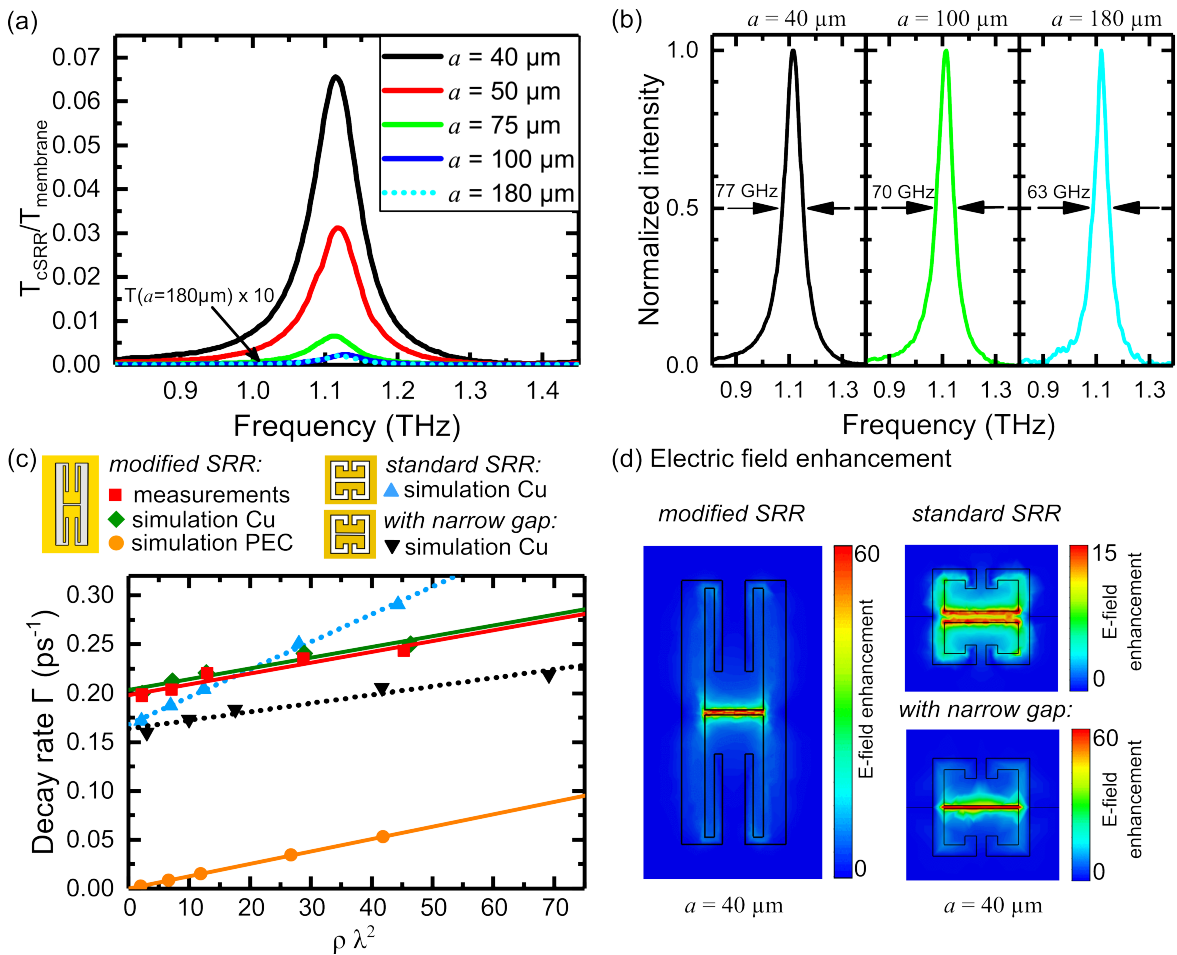

*Figure 2: (a) Transmission intensity spectra of the cSRR array for a = 40 μm (black line), 50 μm (red line), 75 μm (green line), 100 μm (blue line) and 180 μm (cyan dotted line) are shown, normalized by the transmission through the membrane. The transmission of a = 180 μm is multiplied by 10 to be visible on the same scale. (b) Normalized to maximum intensity transmission spectra for lattice constants a = 40 μm, a = 75 μm and a = 180 μm. (c) The decay rate Γ as function of resonator density ρ times the squared wavelength λ for the measured resonators on a Si-membrane with Cu (red squares), simulated Cu resonators with a conductivity of 2.5 x 10$^7$ S/m (green route) and simulated resonators with perfect electric conductor (PEC) material (orange circles). Solid lines in corresponding colors represent linear fitting of the data. The simulated decay rates are also shown for a standard SRR at the same resonance frequency (sketch at top of graph b, outer dimensions are 17 μm x 17 μm with a gap of 2 μm, more details in supporting information Fig. S1)) with Cu (conductivity as above, blue triangles) and for a standard resonator with narrow gap (500 nm, black triangles) with corresponding linear fits (dotted lines). (d) Simulated electric field enhancements for a = 40 μm comparing the standard SRR with normal and narrow gap with the modified SRR.*

The role of the membrane is crucial in order to uncover this fundamental collective effect at large lattice spacings. In our previous study with the same resonator design but a 625 μm thick GaAs substrate,[37] we observed the excitation of surface plasmon polaritons (SPPs), which are tunable via the lattice constant *a*. The SPP modes extend about half of the wavelength into the substrate.[38] At 1.12 THz this corresponds to $\lambda_{vac}/2 \sim 134$ μm in vacuum and $\lambda_{Si}/2 \sim 39$ μm in Si. By placing the resonators on a 10 μm Si-membrane, much thinner than $\lambda_{Si}/2$ at 1.12 THz, we suppress the formation of bound SPP modes at this frequency and shift the SPP modes to much higher frequencies out of range. This allows for an investigation of the pure LC-resonance linewidth as function of resonator density up to very large lattice constants, revealing the underlying collective effect of the resonators.

To understand the contribution of the existing loss mechanism in the system, we simulate the transmission spectra with the commercially available software CST Microwave Studio. Modeling resonators in Cu with a conductivity of 2.5 x $10^7$ S/m perfectly describes our experimental results. The extracted linear slope and the intercept deviate less than 3% from our measurements, and are shown together with the measurements in Fig. 2 (c). The total decay rate $\Gamma$ is the sum of the decay rates of the material losses $\Gamma_{loss}$ and the radiative losses $\Gamma_{rad}$ at the same resonance frequency $f$.[5,18] The radiative losses are proportional to the density of the resonators times the wavelength squared:

$$\Gamma_{tot} = \Gamma_{loss} + \Gamma_{rad}\, \rho\, \lambda^2 \tag{1}$$

For a perfect electrical conductor (PEC) however, the total decay rate is equal to the radiative decay rate since it has no material losses ($\Gamma_{loss} \equiv 0$) by definition. In Fig. 2 (c) we see that the linear decay rate over density for PEC and Cu have a very similar slope but different intercepts. Accordingly, we deduce that the slope and hence the linear dependence is connected to the radiative losses in the system. Instead, the offset between the two linear slopes refers to intrinsic losses of the design and material. Those include Ohmic losses, thin film domain boundaries and other imperfections, which can arise due to the narrow features in the patterning of the thin metallic film with the resonator shape. Employing a superconducting material could eventually minimize the material losses and lead to even higher Q factors and the expense of cooling the sample to cryogenic temperatures.

Similar observations are made when simulating the quality factor for a different resonator shape as function of density. The chosen resonator is a standard complementary split ring resonator[2,6] as sketched in the inset of Fig. 2 (b) at the same resonance frequency as our modified design with outer dimensions of 17 μm x 17 μm and a capacitor gap size of 2 μm (for more detailed dimensions see Fig. S1 in supporting information). If we now compare the intercepts of the linearly dependent decay rates on the density, we observe that the intercept of the simulated Cu standard SRR is slightly lower than for the modified design. This might be due to the fact that the standard design has less narrow and long geometrical restrictions as inductive elements, which could lead to an enhanced material loss effect in our modified SRR. Considering instead the slopes of the linear decay rates of the standard resonator and the modified resonator, we find a steeper slope for the standard SRR than for the modified SRR. Thus, the modified resonator has less collective broadening, i.e. a higher quality factor at high resonator densities. This can be understood by less radiative coupling of the 4 times narrower gap used in the modified design (500 nm) compared to the standard design (2 μm). Such an effect of the capacitor gap size on the linewidth was already described in Ref. [20]. However, in order to keep the resonator frequency the same as in the standard design, the modified resonator design features long inductive elements as the frequency scales as $f \sim 1/\sqrt{LC}$. To further support our reasoning that the superradiant broadening is mainly due to the size of the capacitor gap, we conducted an additional simulation. For this purpose, we keep the standard resonator geometry and only narrow the capacitor gap from 2 μm to 500 nm. This leads, as can be seen from the slope in Fig. 2 (c), also to less superradiant broadening. The density of the simulated fields are kept the same ($a$ = 40 μm, 50 μm, 75 μm, 100 μm and 180 μm) but the frequency of the mode shifts from 1.1 THz to 0.9 THz. The reduced superradiant broadening at high resonator densities for the modified design at the desired frequency clearly overweighs the increased material losses due to enhanced Ohmic losses. Thus, a careful design of the resonator shape allows for improving the quality factor at high resonator densities.

Considering the overall transmission (Fig. 2 (a)) of the different arrays, it naturally decreases as the number of elements per area is decreased in an otherwise full metal plane shielding broadband transmission in the complementary resonators. The transmission for each array size is less than the transmission of the corresponding standard SRR design (see direct comparison of CST simulations in Fig. S2 supporting information). This effect is native to a reduced capacitor gap with less radiative coupling, which in turn leads to the desired high quality factor resonances. The energy, which is not radiative coupled out, is instead concentrated in the capacitor gap, thus leading to high electric fields enhancements favorable to perform, e.g. light-matter interaction studies.[17,51] The electric field distribution of the modified design compared to the standard SRR and the standard SRR with narrow gap are shown in Fig. 2 (d) for $a = 40$ µm. As expected, the electric field enhancement increases from ~15 for the standard SRR to ~60 for the modified design and the narrow gap standard SRR, which have a higher quality factor due to less radiative coupling.

## 2.2 Dipolar mode – substrate related blue shifting of surface plasmon polaritons

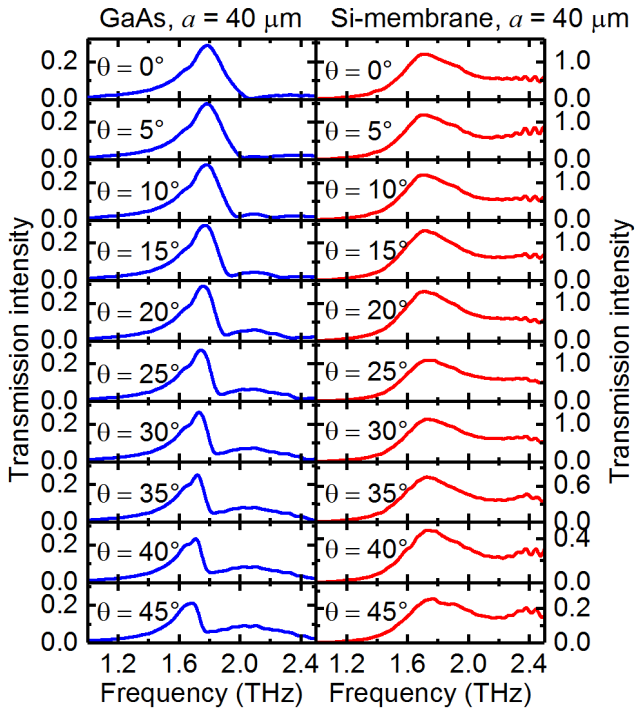

*Figure 3: Left panel: Transmission spectra of the dipolar mode of the cSRR on a 625 µm thick GaAs substrate for different incidence angles θ, showing the dispersive SPP mode. Right panel: Transmission spectra of the dipolar mode of the cSRR on a 10 µm Si-membrane.*

To study the influence of the substrate on the observable modes, we compare a resonator array with lattice constant $a = 40$ µm deposited on a Si-membrane to the *identical* resonator array deposited on a 625 µm thick GaAs substrate that we previously studied in Ref. [37]. Rotating the polarization of the probing THz field by 90° (as indicated in Fig. 1) with respect to the excitation direction of the fundamental LC-mode results in a different resonance mode, the so-called dipolar mode, which is typically at a higher frequency than the LC-mode. For the Si-membrane, the peak frequency of the dipolar mode is at $f = 1.7$ THz. On the thick GaAs substrate, we observe the first order of a surface plasmon polariton mode at $f = 2$ THz, as shown in Figure 2, left column. By rotating the sample around the y-axis (angle θ, see Fig. 1) and probing the dipolar mode polarized along the x-axis, the effective lattice constant to excite the SPP mode is changed. Thus, the frequency of the SPP mode decreases with increasing rotation angle θ (Fig. 3, left column). In Fig. 3, right column, the same measurement, but with the resonators deposited on the Si-membrane, is shown. It is clearly visible that no SPP is excited on the membrane sample. This shows the efficient shift of the bound SPP modes at this frequencies to much higher frequencies by thinning the substrate to less than $\lambda/2$ of the SPP wavelength in the substrate.[38]

**2.3 Photonic crystal modes mediated by cSRRs on a Si-membrane**

Increasing the lattice constant of a cSRR array deposited on a Si-membrane leads to a narrowing of not only the LC-mode, but also the dipolar mode (Fig. 4). We observe a complex and very broad line shape for $a = 40$ µm which evolves into a narrow Lorentzian line shape at $a = 100$ µm with a quality factor as high as $Q = f/\Delta f = 19$ at $f = 1.7$ THz. Additionally, we observe another resonance at $f = 2.2$ THz for a lattice constant of $a = 100$ µm. The frequency of this additional mode depends on the lattice constant. It is shifted to higher frequencies for $a = 75$ µm and lower frequencies for a lattice constant of $a = 180$ µm.

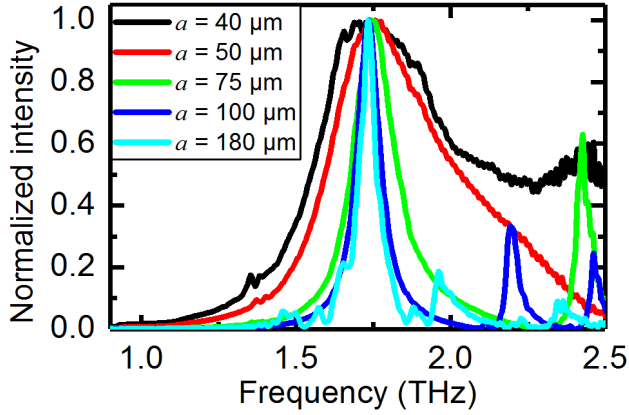

*Figure 4: Transmission spectra of the dipolar mode for five different lattice constants on Si-membranes: a = 40 μm (black line), a = 50 μm (red line), a = 75 μm (green line), a = 100 μm (blue line) and a = 180 μm (cyan line).*

This additional mode, in contrast to the pure dipolar mode in Fig. 4, also exhibits a dispersion as a function of rotation angle θ, as shown in the transmission spectra for a lattice constant of *a* = 100 μm in Fig. 5 (a). The sample is rotated around the y-axis as indicated in Fig. 1. The spectra are taken every 5° and the data is treated with a spline interpolation. The rotation angle θ is converted into the in-plane wave vector *k*. The dipolar mode is at f = 1.7 THz for normal incident, which is slightly shifted to positive angles due to experimental setup conditions. The dispersion is symmetric around a rotation angle corresponding to normal incidence. At large rotation angles, we observe the interaction of the dipolar mode with one of the additional modes. The resonances are anti-crossing, thus are strongly coupling. A CST simulation (Fig. 5 (b)) of the S-Parameter for the same sample rotation in steps of 5° with the same data treatment as for the experimental spectra, is in very good agreement with the experimental data.

Due to the similarity between the measured dispersion and the band structure of a square lattice photonic crystal in the Γ-X direction,[55] we refer to this modes as photonic crystal modes. In typical photonic crystals one engineers a periodically varying refractive index with e.g. air and a material with higher refractive index such as Si,[55] whereas in the present study, a metal layer on top of the Si-membrane is patterned periodically with complementary split ring resonators which also have their own resonance as discussed in the previous sections. Therefore, we ascribe the existence of the observed photonic crystal mode in our samples to the varying metal loss and reflection properties at the resonator position in the thin Si-membrane.

The nature of these modes is further investigated by extracting the electric field distributions simulated with CST. In Fig. 5 (c), the electric field distributions at k = 0 (normal incidence) are shown for the dipolar mode at $f$ = 1.7 THz and the photonic crystal mode at $f$ = 2.12 THz. It is displayed in the y-z-plane, i.e. showing a cross section through the membrane at the center of the resonator, as indicated by the sketched resonator outline. At $f$ = 1.7 THz, the electric field is concentrated at the position of the resonator in the gold layer, as expected for the dipolar resonator mode, and leaking on both sides of the gold layer into the vacuum and the membrane. The field of the photonic crystal modes instead, is concentrated at the interface membrane/vacuum, pushed away from the gold layer. In this case, the 10 μm Si-membrane acts as a slab waveguide and the resonators constitute positions with modified loss and reflectivity such that the electric field is confined in between the resonator positions, leading to an array mode with an angular dependence, in contrast to the SRR resonances. Additionally, (x-y-) in-plane field distributions of the modes can be found in the supporting information (Fig. S3).

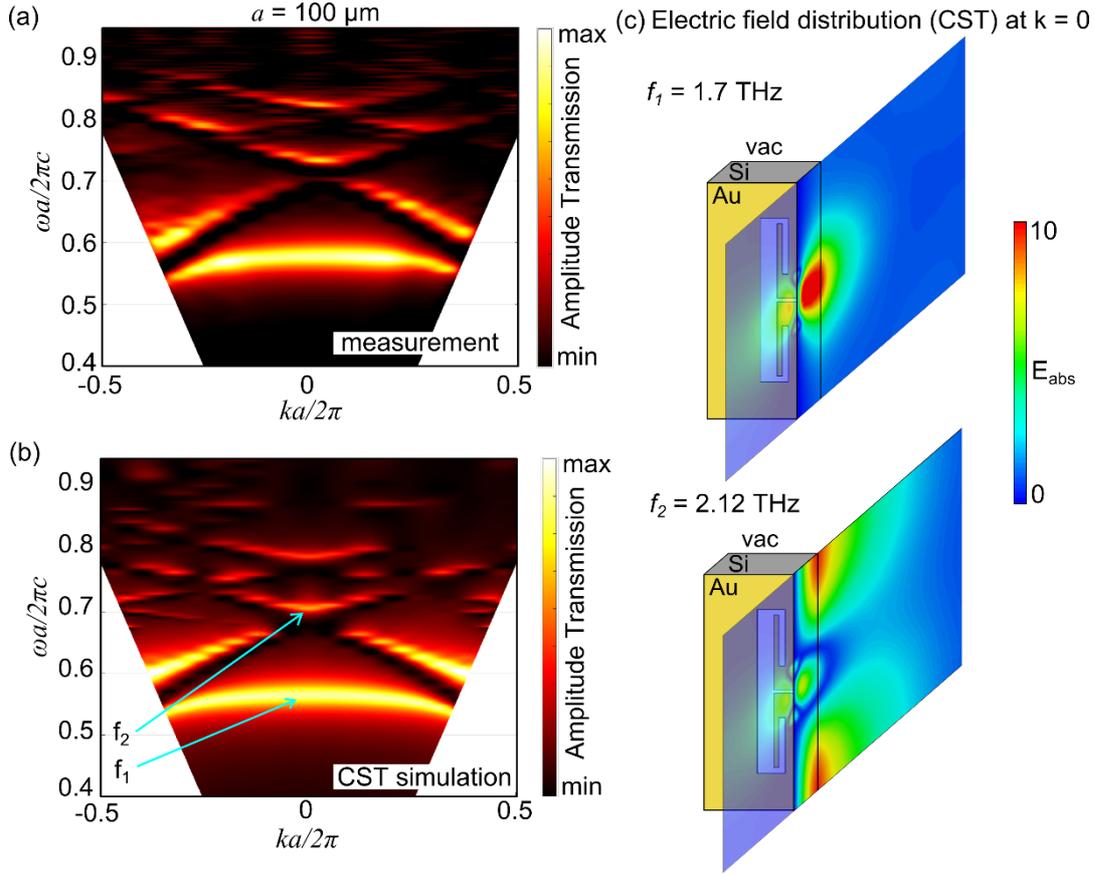

*Figure 5: (a) Measured transmission spectra as function of the in-plane wave vector k are shown. (b) CST simulation of the sample as function of k. (c) Simulated electric field distributions at normal incidence (k = 0) are shown for the dipolar mode at $f_1$ = 1.7 THz, and the first photonic crystal like mode at $f_2$ = 2.12 THz, as indicated by arrows in (b). The position of the resonator in respect to the cross-section view is indicated by an overlaid sketch.*

## 3. Conclusions

In conclusion, we show that the spacing of the meta-atoms has a strong influence on the quality factor of complementary metamaterials. We efficiently shift the surface plasmon polariton modes, which were observed on a thick GaAs substrate in our previous study,[37] out of range to higher frequencies by placing the resonator arrays on a thin Si-membrane. This enables us to observe an underlying fundamental effect: a linear dependence of the quality factor on the density of resonators. This effect was coined "superradiance" by Dicke, originally describing two-level systems radiating in phase and leading to homogenous broadening of the resonance. Here, we relate the density dependent broadening of meta-atom arrays to such collective electric dipole coupling to the incident THz-radiation. This bright mode leads to homogeneous broadening of the underlying resonance. We show that narrow linewidths in the metamaterial limit can be obtained by optimizing the resonant element design for low radiative losses.

Additionally, due to the thin Si-membrane, we allow for the formation of complex-coupled photonic crystal modes, as the periodic array modifies the loss and reflection conditions across the full membrane thickness. These additional photonic crystal modes are at much higher frequencies than the LC-mode and are therefore not interfering with the LC-resonance for all measured lattice constants.

With our study we gain insight into the underlying physical effects which contribute to the quality factor, electric field enhancement and observed transmission resonances of complementary metamaterial arrays on Si-membranes. The results show the strong dependence of the total loss rate on the geometry of the resonators, the spacing between individual meta-atoms, and the effect of the substrate on the formation and the coupling to substrate supported modes.

**Acknowledgements**

We acknowledge financial support from the Swiss National Science Foundation (SNF) through the National Centre of Competence in Research Quantum Science and Technology (NCCR QSIT) and Molecular Ultrafast Science and Technology (NCCR MUST). We also acknowledge financial support from the ERC grant MUSiC.

Table of Contents:

**By placing complementary split ring resonator array with changing density on thin Si-membranes, it is observed that the decay rate depends linearly on the resonator density, thus showing a superradiantly broadened linewidth at high densities.** We design a modified resonator which has low superradiant broadening in the metamaterial limit and thus narrow linewidths.

Keywords: terahertz, metamaterials, complementary split ring resonators, superradiance, photonic crystal

*Janine Keller, Johannes Haase, Felice Appugliese, Zhixin Wang, Shima Rajabali, Gian Lorenzo Paravicini-Bagliani, Curdin Maissen, Giacomo Scalari, Jérôme Faist*

**Superradiantly limited linewidth in complementary THz metamaterials on Si-membranes**

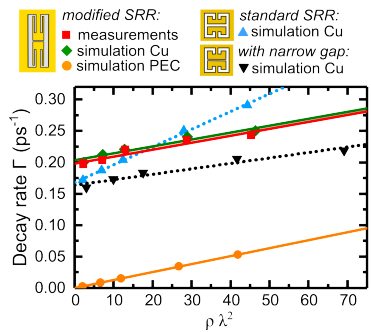

# Supporting Information

**Superradiantly limited linewidth in complementary THz metamaterials on Si-membranes**

*Janine Keller[1]\*, Johannes Haase[2], Felice Appugliese[1], Shima Rajabali[1], Zhixin Wang[1], Gian Lorenzo Paravicini-Bagliani[1], Curdin Maissen[1], Giacomo Scalari[1]\*, Jérôme Faist[1]*

**1. Dimensions of the standard split ring resonator**

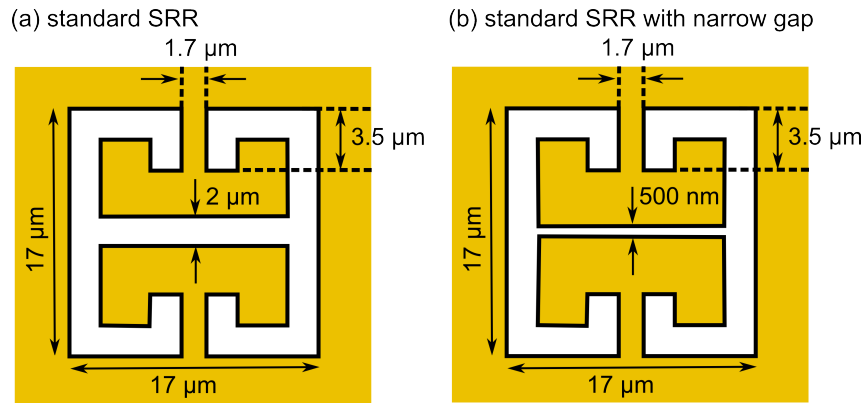

*Figure S1: Standard resonator geometry and standard resonator geometry with narrow gap.*

## 2. CST simulated transmission

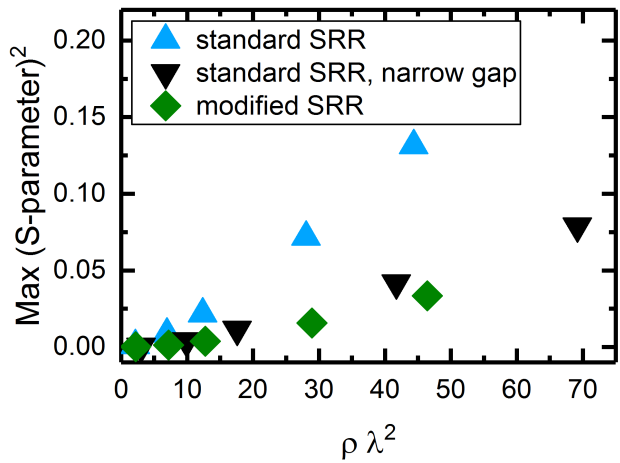

*Figure S2: CST simulated transmission maximum ((S-parameter)$^2$) as function of resonator density times squared wavelength.*

## 4. In-plane (x-y-plane) electric field distributions (to section 2.3)

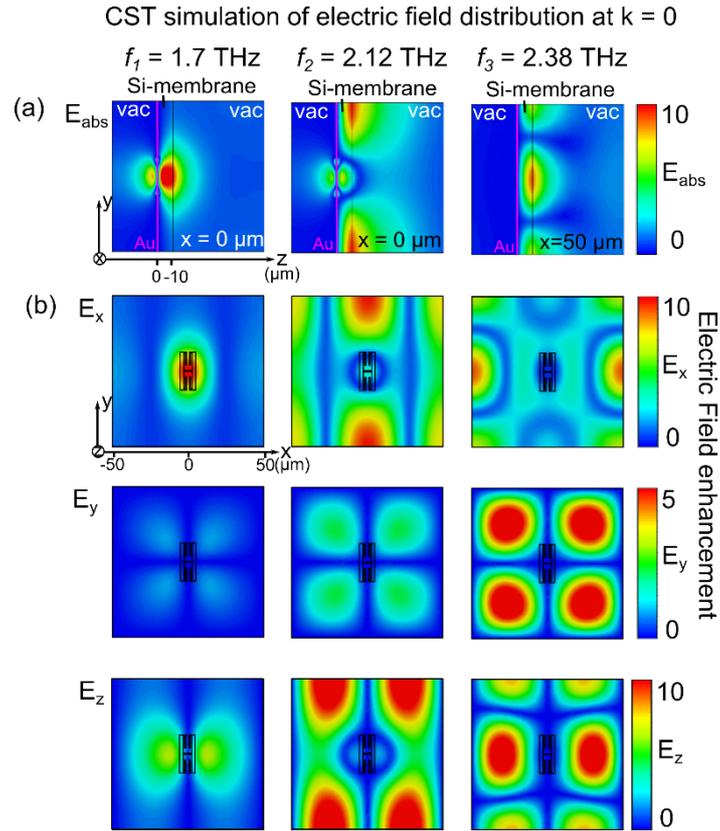

*Figure S3: The electric field distributions at k = 0 for the dipolar mode at $f_1$ =1.7 THz and the photonic crystal modes at $f_2$ = 2.12 THz and $f_3$ = 2.38 THz are shown. In (a) the absolute electric field in the y-z-plane (cross section as in Fig. 5 of the main text) and (b) the electric fields $E_x$, $E_y$ and $E_z$ in-plane (x-y-plane) electric field distribution are displayed.*